\documentclass[aps,prl,twocolumn,amsmath,amssymb,floatfix]{revtex4-1}
\usepackage{tabularx}
\usepackage{bm}
\usepackage{euscript}
\usepackage{graphicx}
\usepackage{color}
\usepackage{amsfonts}
\usepackage{exscale}
\usepackage{amsbsy}
\usepackage{textcomp}
\usepackage{dcolumn}% Align table columns on decimal point
\usepackage{amsmath}
 \usepackage{xfrac}

\def\be{\begin{equation}}
\def\ee{\end{equation}}
\def\ba{\begin{eqnarray}}
\def\ea{\end{eqnarray}}

\def\LBCO{La$_{2-x}$Ba$_x$CuO$_4$}
\def\LBBCO{La$_{1.875}$Ba$_{0.125}$CuO$_4$}
\def\PBSCO{Pb$_{0.55}$Bi$_{1.5}$Sr$_{1.6}$La$_{0.4}$CuO$_{6+x}$}

\def\BSCCO{Bi$_{2}$Sr$_2$CaCuO$_{8+x}$}
\def\YBCO{YBa$_2$Cu$_3$O$_{6+x}$}
\def\YBCOO{YBa$_2$Cu$_3$O$_{6.56}$}

\def\C60{A$_x$C$_{60}$}

\def\HBCO{HgBa$_2$CuO$_{4+x}$}

\begin{document}

%\centerline{ \bf 
\title{Evidence of chiral order in the charge-ordered phase of \LBBCO}
%\bigskip
\author{Hovnatan Karapetyan$^{1,2}$}
\author{Jing Xia$^3$}
\author{M. H\"ucker$^4$}
\author{G. D. Gu$^4$}
\author{J. M. Tranquada$^4$}
%\author{Y. Yoshida$^5$}
%\author{H. Eisaki$^5$}
%\author{N. Bari\v{s}i\'c$^6$}
%\author{M. Greven$^6$}
%\author{D. A. Bonn$^7$}
%\author{W. N. Hardy$^7$}
%\author{ R. Liang$^7$}
\author{M.M. Fejer$^1$}
\author{A. Kapitulnik$^{1,2,5}$}
\affiliation{$^1$Department of Applied Physics, Stanford University, Stanford, CA 94305, USA}
\affiliation{$^2$Stanford Institute for Materials and Energy Sciences, SLAC National Accelerator Laboratory, 2575 Sand Hill Road, Menlo Park, CA 94025, USA}
\affiliation{$^3$Department of Physics and Astronomy, University of California, Irvine, CA 92697-4575, USA}
\affiliation{$^4$Condensed Matter Physics and Materials Science Department, Brookhaven National Laboratory, Upton, NY 11973-5000, USA}
\affiliation{$^5$Department of Physics, Stanford University, Stanford, CA 94305, USA}
%\author{M. Hashimoto$^4$}
%\affiliation{$^4$Advanced Light Source and Materials Sciences Division, Lawrence Berkeley National Laboratory, Berkeley, California 94720, USA}

\begin{abstract}
High resolution polar Kerr effect (PKE) measurements were performed on \LBBCO~single crystals  revealing that a finite Kerr signal is measured below an onset temperature-$T_K$ that coincides with charge ordering transition temperature $T_{CO}$. We further show that the sign of the Kerr signal cannot be trained with magnetic field, is found to be the same of opposite sides of the same crystal, and is odd with respect to strain in the diagonal direction of the unit cell. These observations are consistent with a chiral ``gyrotropic" order above $T_c$ for \LBBCO; similarities to other cuprates suggest that it is a universal property  in the pseudogap regime.
\end{abstract}

\date{\today}

\maketitle

Recently, strong evidence has been mounting that various forms of non-superconducting electronic order occur as general features of the hole-doped high-$T_c$ cuprates (HTSC)  in the pseudogap regime \cite{Norman2004} below some characteristic temperature $T^*$. While the existence of magnetic correlations raises the question of whether  time reversal symmetry ($\mathcal T$) is broken, the appearance of charge order raises the possibility of other broken symmetry states, including inversion ($\mathcal I$) and chirality ($\mathcal C$).  

Time reversal symmetry breaking (TRSB) effects have been predicted to occur within the pseudogap regime \cite{Varma1997,Chakravarty2001}, and have been observed in neutron scattering \cite{Bourges2006,Greven2008,Bourges2012} and magnetic susceptibility \cite{Leridon2009} experiments. Polar Kerr effect  (PKE),  in which the linear polarization of light reflected from a sample at normal incidence is rotated by a Kerr angle $\theta_K$,  is another way to test for TRSB. Indeed, PKE was recently detected below a finite temperature $T_K$, within the pseudogap phase in \YBCO~single crystals and thin films (YBCO)  \cite{KerrYBCO,Kapitulnik2009}, single crystals of  \PBSCO~ (an optimally doped single layer, BSCCO:2201 phase) \cite{He2011}, \LBBCO~(LBCO) \cite{KerrLBCO}, and \HBCO~(underdoped single layer HBCO) \cite{KerrHg1201}. However, a peculiar aspect in the Kerr results, first noted in the data of \YBCO~single crystals  \cite{KerrYBCO,Kapitulnik2009}, and subsequently repeated in measurements on $\sfrac{1}{8}$-doped \LBCO, has been the inability to ``train" the effect by cooling through the transition in an externally applied magnetic field \cite{train}.  While extrinsic effects or incipient magnetism  that orders above room temperature were two possible explanations for these observations, the fact that YBCO single crystals and thin films showed the same $T_K$ and similar magnitude of $\theta_K$  for the same doping level \cite{Kapitulnik2009}, posed an early challenge to the ``contamination scenario."  In addition,  $\mu$Sr \cite{MuSr}, and recent NMR \cite{NMRJulien,NMRKawasaki} studies on YBCO and BSCO \cite{NMRKawasaki}, and HBCO \cite{mounce}, and  neutron scattering measurements on LBCO \cite{tranquada}, find no evidence for magnetic order at any temperature above $T_K$.  Thus, the absence of training effect may point to a new electronic state in the cuprates, which PKE measurements are uniquely suitable to explore.

In this letter we provide further evidence for the uniqueness of the electronic state in \LBBCO.  In addition to {\it i}) the onset of a Kerr signal at a temperature that coincides with charge ordering $T_K\approx T_{CO}$ \cite{tranquada}, and {\it ii}) the inability to train the sign of the Kerr signal with applied magnetic field while cooling the sample through $T_K$ \cite{tranquada}, we also show that {\it iii}) the sign of $\theta_K$  below $T_K$ is the same for reflection on opposite surfaces of the same crystal (although the initial sign can vary among crystals), and {\it iv}) when detected, the change in Kerr signal in response to an applied strain  in the (110) is odd with respect to the response in the (1$\bar{1}$0) directions. The first three properties, further cast doubt on the standard interpretation that our Kerr measurements  identify a simple magnetic-like TRSB, while the simplest way to understand the fourth point is to invoke chirality.  Our results therefore suggest that the Kerr effect in \LBBCO~ arises from a novel ``gyrotropic" order that originates from chiral charge ordering in the pseudogap regime \cite{pavan_prb}. The fact that we find an onset of Kerr signal that coincides with charge ordering in other cuprates further suggest that it is a universal property  in the pseudogap regime \cite{future}.

\begin{figure}[h]
\begin{center}
\includegraphics[width=1.0 \columnwidth]{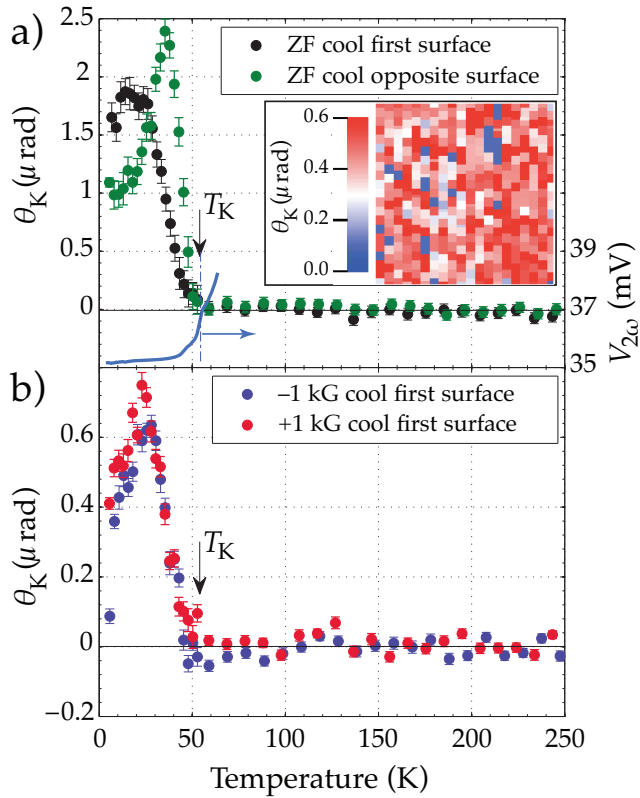}
\end{center}
\vspace{-5mm}
\caption{Kerr angle data of a typical $\sfrac{1}{8}$-doped LBCO~single crystal.  For {\it all} LBCO samples measured $T_K$ coincides with the charge-order transition $T_{CO}\sim 54$~K (see text). (a) ZFW Kerr data taken on two parallel opposite $ab$- surfaces after ZFC. Also plotted is the intensity of the interfering light (i.e. ``second harmonic" ) in the vicinity of the Kerr transition \cite{KerrLBCO}, which clearly show  the anomaly at the first-order structural transition from low-temperature-orthorhombic to low temperature tetragonal $T_{LT}\sim T_{CO}$ \cite{tranquada}. Inset is a scan of 20$\times$20 array of pixels, each 10$\mu$m in size of the local Kerr angle at $T=4$ K in some typical part of a crystal. (b) ZFW data taken on the ``first surface," after  $\pm 1$~kG FC.  } 
\label{fig:lbco}
\end{figure}

The growth technique and crystal properties of the samples used for the data presented in this letter were discussed previously \cite{KerrLBCO}.  PKE measurements were performed using three different zero-area-loop Sagnac interferometers (ZASI)  \cite{Xia2006,Kapitulnik2009}, one which also included a scanning capability. Where possible, birefringence data was inferred from the intensity of the interfering beams (measured in the  ``second harmonic" channel of the ZASI) as explained in ref.~\cite{KerrLBCO}. All Kerr data, whether the sample was cooled in zero field (ZFC) or finite  fields (FC), were recorded in a magnetically shielded environment with residual field $<5$~mG (ZFW) \cite{train}. We identify $T_K$ as the onset of Kerr signal above the noise, and, while the size of the Kerr effect may vary on different points of the samples (and hence is expected to be different in each cooldown), the sign of the effect does not change for a given crystal, as demonstrated in the inset to Fig.~\ref{fig:lbco}a.

To set the stage for our unusual findings, we show in Fig.~\ref{fig:lbco} Kerr angle measurements on a single crystal of \LBBCO~similar to the crystals reported in \cite{KerrLBCO}. In Fig.~\ref{fig:lbco}a, which was measured at zero field after ZFC, we recognize our earlier results \cite{KerrLBCO} in which the Kerr signal is very small (practically zero) above the low-temperature-orthorhombic to low-temperature tetragonal transition (seen also as a sharp drop in the ``second harmonic" - $V_{2\omega}$ signal in that figure, which marks a sharp change in the birefringence of the crystal), which also marks the onset of charge ordering  at $T_{CO}\sim54$~K \cite{tranquada}. Below $T_{CO}$  the Kerr signal increases dramatically peaking at a maximum, (which in \cite{KerrLBCO} was identified as the spin-order transition $T_{SO}$ \cite{tranquada}), and leveling at low temperatures (where in-plane superconductivity is established \cite{KerrLBCO}.)  In Fig.~\ref{fig:lbco}b we plot Kerr data taken on the ``opposite surface" after FC in $\pm 1$~kG, attempting to train the sign of the Kerr signal.  While the peak position of the Kerr signal shifted by about 13~K towards lower temperatures, and the overall size is reduced, the sign of the signal did not change. Since the size of the signal, as well as the peak position have been shown to vary for different samples, as well as different spots on the same sample \cite{KerrLBCO} (see in particular inset of Fig.~\ref{fig:lbco}a), we conclude that the sample cannot be trained with magnetic field.  This is similar to our previous measurements in which we ``failed" to train the LBCO crystals with magnetic fields exceeding 4 tesla \cite{KerrLBCO}. 

Shifting back to Fig.~\ref{fig:lbco}a, we concentrate now on the ``opposite surface" data, noting again a change in the size and peak position of the signal, but with no change to the sign of the Kerr signal.  This is one of the central findings of this paper, which together with the in ability to train the sample with magnetic field prove that the observed Kerr signal is not a result of simple TRSB phenomenon, in which the light couples to an order parameter which is an axial vector, such as the uniform magnetization of a ferromagnet, or a canted component of the magnetization in the case of an  antiferromagnet. In such case,  when the sample is cooled in zero field, the sign of the Kerr signal is expected to vary randomly in sign (and  in magnitude for multiple magnetic domains,) between different thermal cycles, while when cooled slowly through $T_K$ in the presence of an applied magnetic field, the domains are expected to align in the direction of the field, yielding a Kerr signal with a sign commensurate with this alignment. Such unusual behavior can only be explained by a chiral state \cite{VarmaKerr,pavan_prb}.

\begin{figure}[h]
\begin{center}
\includegraphics[width=1.0 \columnwidth]{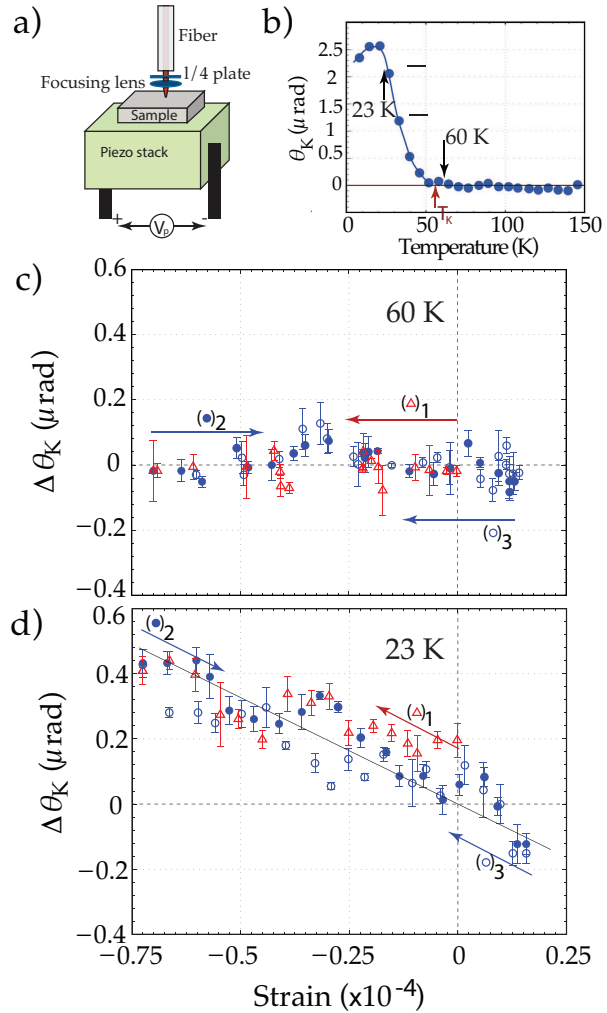}
\end{center}
\vspace{-5mm}
\caption{Kerr angle data of a strained $\sfrac{1}{8}$-doped LBCO~single crystal.  (a) The experimental set-up. (b) The Kerr response as a function of temperature. Here we also mark the two temperatures at which strain-induced changes in Kerr angle are measured. The two horizontal lines indicate the range of change of the Kerr effect due to strain at 23 K. (c)-(d) Kerr angle vs. strain at 60 K and 23 K. Poisitive strain is applied in leg-1, than polarity is reversed in leg-2, and then reversed again in leg-3 (see text). Strain is measured through change in reflectivity (see eqn.~\ref{reflect}). While the Kerr angle did not change within this cycle at 60 K, it did change at 23 K.  } 
\label{fig:lbcopiezo}
\end{figure}

To further probe the nature of the signal we mounted \LBBCO~crystals on the side wall of a piezostack (see Fig.~\ref{fig:lbcopiezo}a for the experimental configuration).   Strains can then be applied by the deformation of the piezo, which is controlled by an applied voltage, as previously used for the study of nematic electronic states  in quantum-Hall systems \cite{Shayegan}, and in iron pnictide superconductors \cite{JHChu}. Using the same type of piezostack, Chu {\it et al.} \cite{JHChu} employed a strain gauge to determine the transmitted strain, finding a maximum strain $\epsilon = \Delta L/L < 10^{-4}$, for samples of thickness of order 0.1 mm and temperatures below $\sim$100 K (here $\Delta L$ is the change in the length of the sample of length $L$ in the direction of the strain). In our experiment we used the relative change in intensity of light coupled back to the fiber after being reflected from the sample  $ \Delta I/I$ to determine the strain. Assuming strong clamping of the crystal to the piezo-stack, we can use $|\Delta L/L|\approx |\Delta t/t|$ where $\Delta t$ is the change in the sample thickness  $t$. Within a simple model of a single gaussian mode of light in the fiber, it can be shown that the efficiency $\eta$  of coupling of light to the fiber due to a defocusing distance $\Delta z$ is $\eta = 1/[1+(\Delta z/2Z_R)^2]$, where $Z_R=\pi w^2/4\lambda$ is the Rayleigh range of coupling ($w = 10 \mu$m is the beam waist and $\lambda =1.55 \mu$m is the wavelength) \cite{Gaussian_Beam}. We estimate an initial efficiency of coupling of light to the fiber at $\eta \sim 0.75  -  0.8$, due to mechanical misalignment and temperature effects.  Since $Z_R \approx 50 \mu$m is much smaller than the focal length of the lens ($\sim 1.5$mm), and $\Delta t \ll \Delta z$ (the initial misalignment is much larger than the strain effect), we expect $\Delta I/I \equiv \Delta \eta/\eta \propto |\Delta t/t|$. Indeed, using simple geometrical optics to estimate the fractional change in sample distance, (i.e. strain) we find
\begin{equation}
\epsilon \approx \frac{\Delta t}{t} \approx \left(\frac{l_s}{l_p}\right)^2\frac{Z_R}{t}\frac{\Delta I}{I}
\label{reflect}
\end{equation}
Here $l_s\approx 2$mm and $l_p \approx 6$mm are the distances of the sample and fiber from the lens respectively, and the sample thickness was $\sim 400 - 500 \mu$m. The above simple estimate gives a maximum strain of $\sim10^{-4}$ in agreement with  Chu {\it et al.} \cite{JHChu}

Fig.~\ref{fig:lbcopiezo} summarizes our strain results on one \LBBCO~crystal. First, in Fig.~\ref{fig:lbcopiezo}b we show the Kerr transition for that sample. Figs.~\ref{fig:lbcopiezo}c,d show the change in Kerr effect as a function of strain. The strain is proportional to the piezo-voltage, as was discussed above, and the full scale change in piezo voltage is calculated from the change in reflected light intensity to be $\sim 10^{-4}$.   In each case the temperature was monitored to be constant within 0.3 K, and the slight hysteresis of the piezo response as measured by the intensity of the reflected light to be $< 4\%$ at 60K and  $<2\%$ at  23 K \cite{JHChu}. Neither influence the Kerr response within the scatter of the data. Fig.~\ref{fig:lbcopiezo}c shows the result at 60 K, just above the Kerr transition, where we cycle the piezo voltage from zero, to $+$150 V, back to $-$50 V and up again to $+$150 V. The Kerr angle is then plotted against the approximate strain values which are calculated from the change in reflectivity using  eqn.~\ref{reflect}.  Here we assign zero stress to zero voltage. While indeed differential thermal contractions of the sample holder, piezo, epoxy, and sample are all different and may cause a bias stress below room temperature, it is expected that such stress is uniform and therefore does not change the behavior we observe when applying a uniaxial stress. Fig.~\ref{fig:lbcopiezo}d shows the result at 23 K, where a strong change in the Kerr angle is observed. Error bars for both figures represent a statistical average. Since at 23 K the sample is structurally tetragonal, and the charge order changes phase by $\pi/2$ every Cu-O layer \cite{tranquada}, any change in Kerr angle that originates from either $c$-axis coupling, or simple anisotropy in the $a$-$b$ plane, would exhibit an extremum at zero strain. The monotonic response of the Kerr effect through zero strain suggests that it is odd under $(110)\rightarrow(1\bar{1}0)$, hence points towards a chiral state.

The inconsistencies of the PKE results with ``simple" TRSB phenomena found in our measurements, have recently prompted several theoretical studies to explain these anomalous results \cite{OrensteinKerr,pavan_prb,VarmaKerr,orensteinmoore,Pershoguba}.  In one set of proposals that requires TRSB, the HTSC is argued to posses a ``magnetoelectric state," in which the Hamiltonian of the system acquires a term $\alpha_{ij}E_i H_j$, where $\alpha_{ij}$ is a rank-2 pseudotensor.  Such a ``magnetoelectric" system also breaks inversion symmetry $\mathcal I$, but preserves the product  $\mathcal T  \mathcal I$ \cite{VarmaKerr,OrensteinKerr}.  Training of the  $\mathcal T $ effect is not possible if only in-plane components  of $\alpha_{ij}$ are present, and in this case opposite sides of the sample will exhibit the same sign of $\theta_K$ \cite{OrensteinKerr}.  In a concrete example, Aji {\it et al.} \cite{VarmaKerr} start from  a so-called ``loop-current state" \cite{Varma1997}, where current loops within a unit-cell lead to two opposing orbital-magnetic moments which order antiferromagnetically at $T^*$ to a magnetoelectric state. While this state should not exhibit a Kerr effect, they propose that at $T_K$, another loop-current state which breaks chirality $\mathcal C $ rather than inversion, but preserves the product $\mathcal T  \mathcal C$ \cite{SunFradkin} appears in addition to the initial state. Such a ``magneto-chiral" state, which is allowed in a multiband description of the cuprates \cite{Emery1987}, will exhibit a finite Kerr effect which cannot be trained by a magnetic field $\textbf{H}||\hat{c}$, and will not change sign on opposite sides of the sample \cite{yanhe}. Indeed, a cascade of transitions was recently identified in ultrasound measurements on \YBCOO~ \cite{Shekhter2012}. 

An alternative approach invokes the fact that reflection from a dissipative medium that lacks a center of inversion, such as a ``gyrotropic" \cite{Bible} material, also results in a non-zero $\theta_K$, even if $\mathcal T$ is preserved \cite{gorkov1992,Bungay1993,Mineev2010,pavan_prb}. A non-local term that appears in the constitutive relations of the material results in an additional term to the dielectric function $\Delta \epsilon_{ab}(\omega,\bm k) = i \gamma_{abc}(\omega)k_c$  ($a,b,c$, are the crystal axes). The gryotropic tensor $\gamma$ is antisymmetric with $\gamma_{abc}(\omega)=-\gamma_{bac}(\omega) \neq 0$ if  inversion  ($\mathcal{I}$) and all mirror symmetries are broken \cite{Bible}. Such a ``handed medium"  can in principle exhibit a Kerr effect  \cite{gorkov1992,Bungay1993,Mineev2010}, however,  to actually obtain a finite $\theta_K$, at least one out of $ \epsilon(\omega)$  and $\gamma(\omega)$ need to be complex and hence, $\epsilon_{ab}(\omega,\bm k)$ to be non-Hermitian, which implies absorption. Surface roughness further enhances the response at optical frequencies \cite{gorkov1992}. Since $\mathcal T$ is not broken, this ``gyrotropic" Kerr effect cannot be trained with magnetic field, and is even under flipping the sample \cite{pavan_prb,orensteinmoore}

Examples of ``gyrotropic" systems include electron analogs of cholesteric liquid crystals, and related systems with chiral density-wave ordering.  Indeed, we have presented above strong evidence that $T_K$ coincides with the onset of charge order. In addition we showed that strain in the (110)-direction, which is the direction that further breaks in-plane mirror symmetries (and hence strengthen the chiral-gyrotropic order) results in an increased Kerr signal. The fact that both sides of the sample show the same sign of the effect are also in line with a chiral-gyrotropic order.  Also, careful symmetry analysis  \cite{pavan_prb} of  Hall and Nernst effect for possible gyrotropic chiral stripe ordering in the cuprates finds a zero-field Nernst effects in \LBCO~ to onset at a temperature $\sim T_K$, consistent with experiment \cite{NernstLBCO}.  

Finally we note that either the magneto-electric scenario \cite{VarmaKerr}, or the gyrotropic charge order \cite{pavan_prb} predict a finite Kerr response due to a chiral state. However, we showed strong evidence that the Kerr effect results from the  striped charge ordering in \LBBCO. Such charge ordering may acquire a definite chirality due to structural effects when the crystal is made as was recently shown by Poccia {\it et al.}  \cite{Poccia} for \BSCCO~single crystals. At the same time, Pershoguba {\it et al.} proposed a combination of loop-current magnetoelectric state with gyrotropy \cite{Pershoguba}. Thus, more experiments as well as theoretical work are needed to fully understand the nature of chirality in this system. Also we note at this point it is not clear what are the consequences of a gyrotropic state for superconductivity, since at least in \LBBCO~ it is argued that such a state can inhibit interplane coupling and hence long-range superconducting coherence \cite{Berg2009}.

\acknowledgments
Stimulating discussions with Zhanybek Alpichshev, Alexander Fried, Pavan Hosur, Steve Kivelson,   Joseph Orenstein, Srinivas Raghu, and Chandra Varma are greatly appreciated. This work was supported by  the Office of Basic Energy Science, Division of Materials Science and Engineering, US. Department of Energy (DOE), at BNL - under contract No: DE-AC02-98CH10886, and at Stanford - under contract No: DE-AC02-76SF00515. Construction of the Sagnac apparatus was partially supported by NSF through Stanford's CPN.


\begin{thebibliography}{99}

\bibitem{Norman2004}
For a recent review see e.g. M. R. Norman, D. P. Pines, and C. Kallin, Adv. Phys. {\bf 54}, 715 (2005).

\bibitem{Varma1997}
C. M. Varma, Phys. Rev. B 55, 14554 (1997); Phys. Rev. Lett. {\bf 83}, 3538 (1999).

\bibitem{Chakravarty2001}
S. Chakravarty, R.B. Laughlin, D.K. Morr, and C. Nayak, Phys. Rev. B63, 094503 (2001).

\bibitem{Bourges2006}
B. Fauqu\'e, Y. Sidis, V. Hinkov, S. Pailhes, C. T. Lin, X. Chaud, and P. Bourges, Phys. Rev. Lett. 96, 197001 (2006); H.A. Mook, Y. Sidis, B. Fauque, V. Baledent, and P. Bourges, Phys. Rev. B {\bf 78}, 020506 (2008).

\bibitem{Greven2008}
Y. Li, V. Baledent, N. Barisic, Y. Cho, B. Fauque, Y. Sidis, G. Yu, X. Zhao, P. Bourges, and M. Greven, Nature (London) {\bf 455}, 372 (2008).

\bibitem{Bourges2012}
S. De Almeida-Didry, Y. Sidis, V. Baledent, F. Giovannelli, I. Monot-Laffez, P. Bourges, preprint: arXiv:1207.1038.

\bibitem{Leridon2009}
B. Leridon, P. Monod, and D. Colson,  Europ. Phys. Lett. {\bf 87} 17011 (2009).

\bibitem{KerrYBCO}
Jing Xia, E. Schemm, G. Deutscher, S. A. Kivelson, D. A. Bonn, W. N. Hardy, R. Liang, W. Siemons, G. Koster, M. M. Fejer, and A. Kapitulnik, Phys. Rev. Lett. {\bf 100}, 127002 (2008). 

\bibitem{Kapitulnik2009} 
A. Kapitulnik, Jing Xia, E. Schemm and A.Palevski, New J. Phys. {\bf 11},  055060 (2009).

\bibitem{He2011}
Rui-Hua He, M. Hashimoto, H. Karapetyan, J. D. Koralek, J. P. Hinton, J. P. Testaud, V. Nathan, Y. Yoshida, Hong Yao, K. Tanaka, W. Meevasana, R. G. Moore, D. H. Lu, S.-K. Mo, M. Ishikado, H. Eisaki, Z. Hussain, T. P. Devereaux, S. A. Kivelson, J. Orenstein, A. Kapitulnik, and Z.-X. Shen, Science {\bf 331}, 1579 (2011).

\bibitem{KerrLBCO}
H. Karapetyan, M. H\"ucker, G. D. Gu, J. M. Tranquada, M. M. Fejer, J. Xia, and A. Kapitulnik, Phys. Rev. Lett. {\bf 109}, 147001 (2012).

\bibitem{KerrHg1201}
In addition to previously published results on YBCO \cite{KerrYBCO}, BSCO \cite{He2011}, and LBCO \cite{KerrLBCO}, we recently measured single-layer Hg-HTSC single crystals which show similar results. H. Karapetyan, N. Barisic, Y. Cho, M. Greven and A. Kapitulnik, to be published.

\bibitem{train}
Throughout this paper, the protocol for magnetic field training includes the following steps: first the sample is cooled from ~250K in the specified magnetic field {\bf H},   with {bf H} $|| \hat{c}$. Then, the sample is moved to a magnetically shielded environment with residual field $< 5$~G.  Kerr effect was then measured with increasing temperature.

 \bibitem{MuSr}
J.E. Sonier, V. Pacradouni, S.A. Sabok-Sayr, W.N. Hardy, D.A. Bonn, R. Liang, H.A. Mook, Phys. Rev. Lett. {\bf 103}, 167002 (2009).

\bibitem{NMRJulien}
T. Wu, H. Mayaffre, S. Kramer, M. Horvatic, C. Berthier, W. Hardy, R. Liang, D. Bonn, and M.-H. Julien, Nature {\bf 477}, 191 (2011).

\bibitem{NMRKawasaki}
S. Kawasaki, C. Lin, P. L. Kuhns, A. P. Reyes, and G.-q. Zheng, Phys. Rev. Lett. {\bf 105}, 137002 (2010); G.-q. Zheng, S. Kawasaki, C. Lin, P. L. Kuhns, A. P. Reyes, and  J Supercond. Nov. Magn. {\bf 25}, 1249 (2012).

\bibitem{mounce}
A.M. Mounce, Sangwon Oh, Jeongseop A. Lee, W.P. Halperin, A.P. Reyes, P.L. Kuhns, M.K. Chan, C. Dorow, L. Ji, D. Xia, X. Zhao, M. Greven, arXiv:1304.6415.

\bibitem{tranquada}
J.M. Tranquada, G.D. Gu, M. H\"ucker, Q.J ie, H.-J. Kang, R. Klingeler, Q. Li, N. Tristan, J. S. Wen, G. Y. Xu, Z. J. Xu, J. Zhou, and M. v. Zimmermann, Phys. Rev. B {\bf 78}, 174529 (2008). 

\bibitem{VarmaKerr}
V. Aji, Y. He, C. M. Varma, Phys. Rev. B {\bf 87}, 174518 (2013).

\bibitem{pavan_prb}
Pavan Hosur, A. Kapitulnik, S.A. Kivelson, J. Orenstein, S. Raghu, Phys. Rev. B {\bf 87}, 115116 (2013).

\bibitem{future}
See ref.~\cite{KerrHg1201}, and J. Xia, H. Karapetyan, A. Kapitulnik, in talk given at conference``Quantum in Complex Matter," Ischia, Italy, June 27-30, 2013, to be published.

\bibitem{Xia2006}
Jing Xia, P. T. Beyersdorf, M. M. Fejer, A. Kapitulnik, Appl. Phys. Lett. 89,  062508 (2006).

\bibitem{Shayegan}
M. Shayegan, K. Karrai, Y. P. Shkolnikov, K. Vakili, E. P. De Poortere, and S. Manus, Appl. Phys. Lett. {\bf 83}, 5235 (2003).

\bibitem{JHChu}
Jiun-Haw Chu,Hsueh-Hui Kuo, James G. Analytis, and Ian R. Fisher, Science {\bf 337}, 710 (2012).

\bibitem{Gaussian_Beam}
A.E. Siegman, ``Lasers" University Science Books, 1986, (ISBN: 0935702113, 9780935702118), Ch. 17.

\bibitem{OrensteinKerr}
J. Orenstein, Phys. Rev. Lett. {\bf 107}, 067002 (2011).

\bibitem{orensteinmoore}
J. Orenstein, Joel E. Moore, Phys. Rev. B {\bf 87}, 165110 (2013).

\bibitem{Pershoguba}
S. S. Pershoguba, K. Kechedzhi, and V. M. Yakovenko, Phys. Rev. Lett. {\bf 111}, 047005 (2013).

\bibitem{SunFradkin}
K. Sun and E. Fradkin, Phys. Rev. B {\bf 78}, 245122 (2008).

\bibitem{Emery1987}
V.J. Emery, Phys. Rev. Lett. 58, 2794 (1987).

\bibitem{yanhe}
Yan He, P. A. Lee, C. M. Varma,  arXiv:1305.7295.

\bibitem{Shekhter2012}
A. Shekhter, A. Migliori, J. B. Betts, F. F. Balakirev, R. D. McDonald, S. C. Riggs, B. J. Ramshaw, R. Liang, W. N. Hardy, and D. A. Bonn, Nature {\bf 498}, 75 (2013).

\bibitem{Bible}
L. Landau, E. Lifshitz, and L. Pitaevskii, Electrodynamics of Continuous Media (Pergamon Press, New York, 1984).

\bibitem{gorkov1992}
B. Arfi and L.P. Gor'kov, Phys. Rev. B 46, 9163 (1992).

\bibitem{Bungay1993}
A. R. Bungay, Y. P. Svirko, and N. I. Zheludev, Phys. Rev. B {\bf 47}, 11730 (1993).

\bibitem{Mineev2010}
V. P. Mineev and Y. Yoshioka, Phys. Rev. B {\bf 81}, 094525 (2010).

\bibitem{NernstLBCO}
Lu Li, N. Alidoust, J. M. Tranquada, G. D. Gu, and N. P. Ong, Phys. Rev. Lett. {\bf 107}, 277001 (2011).

\bibitem{Poccia}
N. Poccia, G. Campi, M. Fratini, A. Ricci, N. L. Saini, and A. Bianconi, Phys. Rev. B {\bf 84}, 100504(R) (2011).

\bibitem{Berg2009}
E. Berg, E. Fradkin, and S. A. Kivelson, Phys. Rev. B 79, 064515 (2009). 

\end{thebibliography}
\end{document}